\documentclass[12pt]{iopart}
\usepackage{graphicx}

\usepackage{iopams}  
\usepackage{color}
\begin{document}

\title[]{A new data analysis framework for the search of continuous gravitational wave signals}

\author{O. J. Piccinni$^{1,2}$, S. Frasca$^{1,2}$, P. Astone$^{2}$, S. D'Antonio$^{3}$,  G. Intini$^{1,2}$, P. Leaci$^{1,2}$, S. Mastrogiovanni$^{1,2}$, A. Miller$^{1,4}$, C. Palomba$^{2}$,  A. Singhal$^{2}$}

\address{$^1$Universita di Roma 'La Sapienza', I-00185 Roma, Italy}
\address{$^2$INFN, Sezione di Roma, I-00185 Roma, Italy}
\address{$^3$INFN, Sezione di Roma Tor Vergata, I-00133 Roma, Italy}
\address{$^4$University of Florida, Gainesville, FL 32611, USA}

\ead{ornella.juliana.piccinni@roma1.infn.it}
\vspace{10pt}
\begin{indented}
\item[] 25.07.2018
\end{indented}

\begin{abstract}
Continuous gravitational wave signals, like those expected by asymmetric spinning neutron stars, are among the most promising targets for LIGO and Virgo detectors. 
The development of fast and robust data analysis methods is crucial to increase the chances of a detection.
We have developed a new and flexible general data analysis framework for the search of this kind of signals, which allows to reduce the computational cost of the analysis by about two orders of magnitude with respect to current procedures. This can correspond, at fixed computing cost, to a sensitivity gain of up to 10\%-20\%, depending on the search parameter space. Some possible applications are discussed, with a particular focus on a directed search  for sources in the Galactic center. Validation through the injection of artificial signals in the data of Advanced LIGO first observational science run is also shown. 
\end{abstract}

%
\vspace{2pc}

\section{Introduction}
\label{sec:intro}
The era of gravitational wave (GW) astronomy has begun on September $14^{th}$ 2015 with the detection of the first gravitational wave event (GW150914) \cite{Abbott2016}. All the GW signals detected so far have been emitted by the coalescence of black holes or neutron stars (NSs) in a binary system 
\cite{Abbott2016,Abbott2017n,Abbott2017a,Abbott2017k,Abbott2017,Abbott2016f,Abbott2016a}. 
Due to their weakness, continuous gravitational waves (CWs) have not been detected yet,  and are subject of intense research within the LIGO and Virgo collaboration.
CWs are emitted by quickly rotating NSs, either isolated or in binary systems, and are characterized by a time-varying quadrupole deformation due to an asymmetry in their mass distribution. Several different mechanisms have been proposed that can produce this asymmetry, such as elastic stresses, strong internal magnetic fields not aligned to the star rotation axis, free precession around the star rotation axis, excitation of long-lasting r-mode oscillations and the accretion of matter from a companion star, e.g. in Low-Mass X-ray Binaries (LMXBs) \cite{Lasky2015,Glampe}. These signals are nearly monochromatic with a frequency proportional to the star spin frequency, with a typical strain amplitude much weaker compared to those of coalescing binary systems, and with a duration longer than the observational time (of the order of months or years). 
The general way to search for CWs depends on how much about the source is known, and full coherent methods (like in \emph{targeted} or \emph{directed} searches) \cite{Astone2010, Astone2014c, Mastrogiovanni2017} or semi-coherent methods (e.g. in \emph{directed} or \emph{all-sky} searches) \cite{GalCentEaH,Dergachev2010, Astone2014a} can be applied.
In the case of directed and all-sky searches a hierarchical follow-up of the most interesting candidates is typically done, in order to approach the sensitivity of a fully coherent search. Although to date no direct CW detection has been claimed yet, remarkable upper limits on the CW signal strength have been obtained \cite{Abadie2011,Aasi2014,Abbott2017o,Abbott2017b_narrow,Aasi2015,Abbott2017Sco2,Aasi2014a,Aasi2016b,Abbott2017w}.
For a review of the latest observational results and methods of CW searches see  \cite{Riles2017,Palomba2017}.
In this paper we present a novel data analysis framework to accomplish CW searches, which provides a gain in sensitivity at fixed computational cost, and in robustness with respect to source parameter uncertainties and instrumental disturbances.
The paper is organized as follows.
In Sec. \ref{sec:DAforCW} we summarize the main features of CW signals.  In Sec. \ref{sec:method} we describe the new data analysis framework and discuss the improvements with respect to current methods used in the LIGO-Virgo collaboration.
In Sec. \ref{sec:injections} we validate the method by injecting artificial CW signals in the data of the first science run of Advanced LIGO (O1 - from Nov. 2015 to Jan. 2016). Several possible applications are described in Sec. \ref{sec:applica}.
 Section \ref{sec:conclusion} is devoted to present conclusions and future perspectives. Some more technical details are shown in the appendices.

\section{The signal}\label{sec:DAforCW}
The expected strain amplitude at the detector, for a CW signal emitted by a non-axisymmetric neutron star, steadily spinning around one of its principal axis of inertia is given by the real part of

\begin{equation} \label{h(t)}
h(t)=H_0\left[H_{+}A^{+}(t)+H_{\times}A^{\times}(t)\right]\mathrm{e}^{i\Phi(t)},
\end{equation}
with  $\Phi(t)=\omega(t)t+\Phi_0$ where $\omega(t)$ is the CW signal angular frequency\footnote{In the case of a NS rotating around one of its principal axes of inertia $\omega(t)=2 \pi f_0(t)= 4 \pi f_{rot}(t)$, where $f_{rot}(t)$ is the star rotational frequency, while $f_0(t)$ is the GW emitted frequency.}, $\Phi_0$ is the phase at the reference time $t_0$, $H_0$ is the maximum signal strain \cite{Astone2010}. 
The complex amplitudes $H_{+} $ and $H_{\times}$ are given, respectively, by: 

\begin{equation}
H_{+} = \frac{ \cos{2 \psi} - i \eta \sin{2 \psi}}{\sqrt{1+\eta^2}}
\end{equation}
\begin{equation}
H_{\times} = \frac{ \sin{2 \psi} + i \eta \cos{2 \psi}}{\sqrt{1+\eta^2}},
\end{equation}
where $\psi$ is the polarization angle. 
The parameter $\eta$  indicates the degree of polarization of the CW and takes values in the range $[-1, 1]$ ($\eta=0$  for a linearly polarized wave, $\eta=\pm1$ for a circularly polarized wave). 
The functions $A^{+}(t)$, $A^{\times}(t)$ encode the detector response to a CW and are equivalent to the standard beam pattern functions $F_{+}(t,\psi)$ and $F_{\times}(t,\psi)$ defined in \cite{Jaranowski1998} for $\psi=0$, where $A^{+}(t)\equiv F_{+}(t)$ and $A^{\times}(t)\equiv F_{\times}(t)$ (see  \cite{Abadie2011}). They depend on the source position, the detector location, orientation  and  sidereal motion. This last effect produces a splitting of the signal power among  five angular frequencies $\omega_0$, $\omega_0\pm\Omega_{\oplus} $ and $\omega_0\pm2\Omega_{\oplus} $, where $\Omega_{\oplus} $ is the Earth sidereal angular frequency. 

Due to the Earth motion, a CW signal arrives at the detector with a frequency modulation (Doppler effect), such that the received signal frequency $f(t)$ is related to the emitted frequency $f_0(t)$ by
\begin{equation}\label{eqn:Doppler}
f(t)=\frac{1}{2\pi}\frac{d\Phi(t)}{dt}=f_0(t)\left(1+\frac{\vec{v}(t)\cdot\widehat{n}}{c}\right),
\end{equation}
where $\vec{v}=\vec{v}_{orb}+\vec{v}_{rot}$ is the detector velocity, sum of the Earth's orbital and rotational velocity, while $\widehat{n}$ is the unit vector pointing to the source position, both expressed in the  Solar System Barycenter (SSB) reference frame\footnote{This relation is valid in the non-relativistic approximation $v/c \ll 1$.}.
If the source is in a binary system a further Doppler modulation, due to the orbital motion, is present \cite{Leaci2017}. In this paper, for simplicity, we focus on isolated sources only.

The emitted signal frequency, $f_0(t)$, slowly decreases with time due to the rotational energy loss of the star, consequent to the emission of both EM and GW radiation. This effect, called spin-down, can be described by a Taylor series expansion:
\begin{equation}\label{eqn:spindown}
f_0(t)=f_0 + \dot{f_0}(t-t_0)+\frac{\ddot{f_0}}{2}(t-t_0)^2 + \ldots,
\end{equation}
where $[\dot{f_0},\ddot{f_0},\ldots]$ are the spin-down parameters. There are also other relativistic effects, namely the Einstein and the Shapiro delays,  which effects in some cases are not negligible \cite{ShapiroEinsteindelay}.
All these phase modulations can prevent a possible detection if not properly taken into account.

\section{The method}\label{sec:method}
The motivation for this work is to have a proper infrastructure,  based on an efficient data cleaning, organization and management, where frequency sub-bands and/or time sub-periods can be quickly and easily  extracted and then given in input to the analysis pipeline, as this is very relevant for all possible searches, in particular those listed in Sec. \ref{sec:intro}, which are the main target of our work. Hence,  we introduce a new data framework, which we refer to as ``Band Sampled Data" collection (BSD), that consists of band-limited, down-sampled time series, which we clean accordingly to the procedure described in \ref{sec:cleaning}.  
We do not exclude, however, a possible use of this new framework for other searches, like the search for transients signals from supernova explosions or Compact Binary Coalescence signals (see \ref{sec:transients}), where the application of an additional cleaning procedure based on glitch removal and time/frequency sub-bands quick extractions, might be helpful in order to prepare the data for further analyses, bases e.g. on matched filter or excess power.
We stress that a good and quick data management is equivalent to an increase in sensitivity, as we are typically limited by the available computing power. In other terms, at fixed computing cost, we can explore with a better sensitivity a broader parameter space. This can even make the difference between a missed signal and a detection. 
A similar method, exclusively used for the search of signals from known pulsars is reported in \cite{Dupuis2005}. Comparing to \cite{Dupuis2005} the main difference is that our method is more general, since we can correct the data as a final step, allowing to perform more than a targeted search, including the possibility to correct the data using parameters which are slightly different from those of the targeted pulsar. 
Another recent method to speed up CW searches has been proposed in \cite{Keitel2018}. This method, specifically developed for transients CW signals, lowers the computational time required for the search by exploiting the use of GPUs and python libraries.

\begin{figure}[h!]
\centering
\includegraphics[scale=0.55,trim=0.5cm 6cm 0.5cm 4cm,clip]{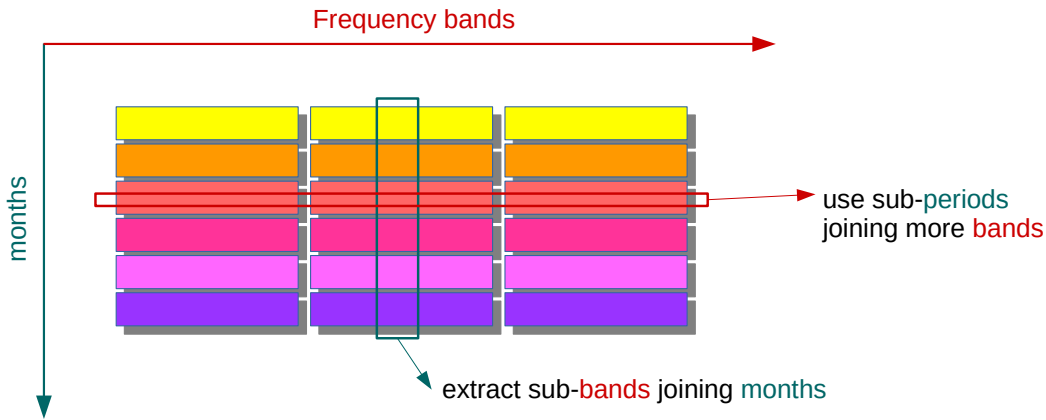}
\caption{Time versus frequency representation of BSD data, where each block represents a 10 Hz and 1 month of data (one BSD file). Using BSD libraries it is possible to extract frequency sub-bands (less than 10 Hz) or time sub-periods (less than 1 month). As well,  it is possible to select larger frequency bands (more than 10 Hz), grouping the 10 Hz pieces, and/or larger-periods (more than 1 month), grouping the 1-month data.}
\label{fig:DBofDBs}
\end{figure}

The core of the procedure used to construct the BSD database and manage it, for next analyses, is described in the following.
We start from a collection of overlapped short Fourier Transforms (computed through the Fast-Fourier-Transform algorithm, here referred as FFTs), like the ``Short FFT Data Base" (SFDB)\footnote{where a preliminary cleaning procedure for strong time glitches, over the full frequency band, has been applied}  \cite{Astone2005}, used for all-sky CW analysis in the Virgo collaboration. From the SFDB files we extract a band of 10 Hz (which is a tunable parameter), compute the Inverse Fourier Transform (IFFT) and thus create a smart representation, as detailed below, of the analytic signal of the data, sub-sampled by a factor $\frac{2F_{max}}{10}$, where $F_{max}$ is the maximum frequency of the FFTs we are using (e.g. 1024 Hz).
Since the FFTs  in the SFDB files are overlapped in time by one half, we eliminate the overlapping effect\footnote{overlapping data before constructing FFTs is important to remove edge effects when the same data are used to reconstruct the time domain series. Edge effects might be due to the windowing procedure, which is needed to reduce the energy spread of the quasi-sinusoidal signal in the frequency domain}, by removing the first and the last quarter of data, hence keeping only the central part of the time series. This is repeated for all the FFTs spanning one month of data, and in the end one file for each month is produced. The procedure is then repeated for all the 10 Hz sub-bands.
In the end, we have, for each detector, a collection of 10 Hz/1 month cleaned complex time domain data, which we typically refer to as the ``BSD files" and which forms a database of sub-databases.
It is important to add that the associated BSD library of functions, allow to produce intermediate data information, which can be used, if needed, for the next analyses. 
Figure \ref{fig:DBofDBs} shows the basic principle of frequency sub-bands or time sub-periods extraction.
We show the frequency representation of BSD database versus time, where each frequency block represents a 10 Hz/1 month of data (a BSD file). Larger frequency bands, grouping the 10 Hz pieces, and/or larger time periods, grouping the 1-month data, can be selected too. The BSD library permits to elaborate the files as required and rapidly produce one single instance of the data, ready for the next step of the analysis.
This flexibility allows, for example, to create a set of FFTs with a duration (and hence a frequency resolution)  optimized for the search. The BSD framework includes also detailed information, stored as a header in every file, which is needed in several steps of the analysis.
The information is, for example, the beginning time of each data set and  the position and velocity of the detector in a chosen reference frame (we use the solar system barycenter), which is sometimes used to remove the Doppler effect in the case of CW searches.
 
\subsection{The BSD creation and basic usage}\label{sec:creation}
\begin{figure}[h!]
\begin{center}
\hspace*{0cm}
\includegraphics[width=\linewidth,trim=0.2cm 0.1cm 0.1cm 0.1cm,clip]{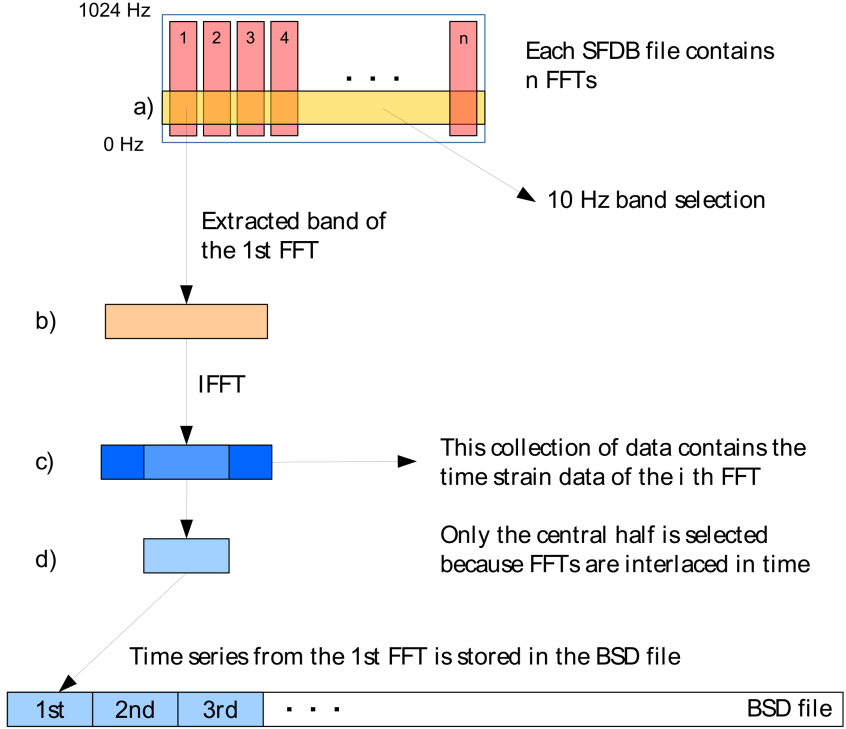}
\caption{(color online) Detailed BSD preparation scheme. On the vertical axis we represent the frequency, from 0 to 1024 Hz, while on the horizontal axis we represent the time, from the beginning to the end of a run. Each FFT appears as a vertical box (pink, labeled from 1 to n). We have then indicated with the yellow box (horizontal, a)) the 10 Hz we want to extract. The extraction is represented for the first FFT and indicated with the orange box (b)). Inverse FFT and selection of the central part of the time series (light blue, central part of the c) block) are then indicated. Finally, the data obtained by repeating this process for all the $n$ FFTs are attached to reconstruct the final 10 Hz/1 month BSD file.}
\label{fig:schema}
\end{center}
\end{figure}

In this section, we give additional information about the basic procedures used to create the sub-sampled and complex time domain BSD data (organized in blocks of 10 Hz/1 month).
First of all, we remind that the starting point is a database of overlapped by half in time FFTs. Given the set of FFTs, the BSD construction is computationally very cheap and fast. In general it is possible to create the BSD data directly from the detector raw data, but practically we started from the SFDB. As an example, the total creation time for a band of 1024 Hz, four months of data and two detectors,  is only a few core-hours.
The needed storage capacity is also very small, of the order of 260 GB.
Figure \ref{fig:schema} shows a flow diagram of the BSD production scheme.
On the vertical axis we represent the frequency, from 0 to 1024 Hz in the considered case, while on the horizontal axis we represent the time, from the beginning to the end of a run. So each FFT appears as a (vertical, labeled from 1 to n) box. We have then indicated with an horizontal a) labeled box, the 10 Hz we want to extract.
The extraction, and hence the sub-sampling, is done on each FFT and is represented by the b) box for the first FFT.  Given that in the original FFT database we have stored only the positive part of each FFT, the construction of the classical analytic time series would require to add zeros to the negative frequency part of each FFT before performing the inverse FFT (IFFT).
In this new approach, detailed in \ref{sec:ana_signal}, we are able to recover the time domain data by using only the positive half of each FFT. This brings to the construction of a complex time series, which we will call \emph{reduced-analytic} time series (in figure represented by the c) block).
We then select only the central part of each reconstructed time series (central part of the c) block, represented by the d) block). A further cleaning procedure is then applied (see  \ref{sec:cleaning}).
At this point, Doppler and spin down removal, that is signal demodulation based on the heterodyne method, can be done.
This is an important part of the procedure, which takes advantages of having a sub-sampled set of data and which is described in the next section, Sec. \ref{sec:correction}.
Finally, the data obtained by repeating this process for all the FFTs are attached to reconstruct the final 10 Hz/1 month file. The new sampling time is $\frac{1}{\Delta f_{BSD} }$, where $\Delta f_{BSD}=10$ Hz.
The procedure described above, is repeated for all the 10 Hz sub-bands we need to cover, which is roughly a hundred times for the band [10-1024] Hz, and is organized in order to work in parallel for more bands, in order to reduce the overall time needed to load the files. The only limit arises from the available RAM.
We stress that the proposed procedure, compared to the construction of the analytic signal,  has the advantage of reducing the computing time (half data to be handled) in the data processing, while saving the full information needed to analyze the data or even reconstruct the full time series, if needed. Details about the difference between the classical analytic signal and its reduced form are described in \ref{sec:ana_signal}. We notice however that in CW typical analyses we don't need to reconstruct the original time series, as the analysis are mainly based on the spectral characteristics of the data. 
In addition, BSD creation implies also the storage of relevant parameters, in the so-called ``auxiliary data structure'', like those needed for further data processing, e.g. the Doppler removal, for which the position and the velocity of the detectors in the SSB are needed.

\begin{figure}[h!]
\begin{center}
\hspace*{0cm}
\includegraphics[width=0.7\linewidth]{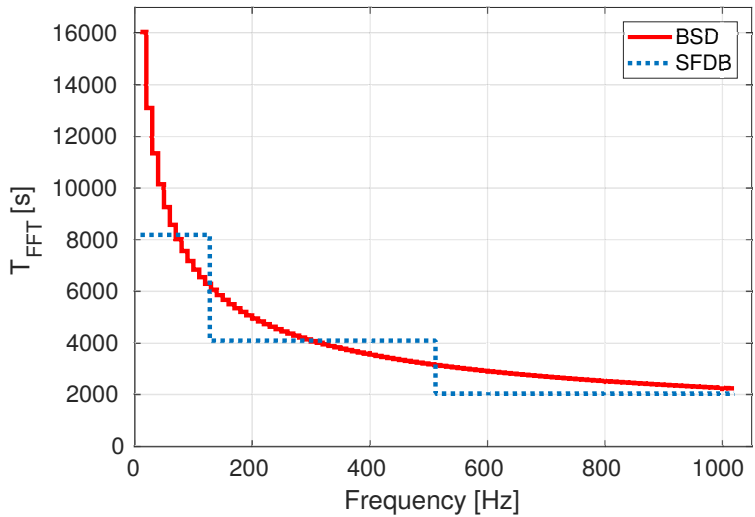}
\caption{Here we show the FFT duration that can be used in the two cases (the actual approach, based on four FFT sets, called the SFDB, and the new BSD approach), as a function of the frequency. We observe that in the lowest frequency region the improvement is of the order of $\sqrt[4]{2}$ since we can use FFTs two times longer than those used in the SFDBs.} 
\label{fig:FFTlength}
\end{center}
\end{figure}

We can now show the sensitivity improvement obtained by using this approach in LIGO-Virgo semi-coherent searches, where the typical compromise between the limited computing power and the unknown source parameters is the construction of FFT databases, using different FFT lengths for searches done up to a given maximum frequency. As will be clear in the following the sensitivity of a semi-coherent search is strictly related to the FFT length used in the coherent step of the analysis. For example, as described in \cite{Astone2005}, the FFT duration $T_{FFT}$ is chosen in such a way that a GW signal frequency is shifted (due to the Doppler modulation) by no more than one frequency bin (which is given by $\frac{1}{T_{FFT}}$) during the FFT duration.  When four databases are used, that is the choice done in \cite{Aasi2016b}, they cover respectively the frequency band $[10-128]$ Hz (using $T_{FFT}=8192\, \mathrm{s}$), $[128-512]$ Hz (using $T_{FFT}=4096\, \mathrm{s}$), $[512-1024]$ Hz (using $T_{FFT}=2048\, \mathrm{s}$) and $[1024-2048]$ Hz (using $T_{FFT}=1024\, \mathrm{s}$). It is hence easy to understand that many frequencies are penalized by these choices, e.g. the  $T_{FFT}=8192\, \mathrm{s}$ choice is optimized for searches at 64 Hz,  but could be more for searches at 20 Hz, while it is higher than the optimal value at 128 Hz.  The huge flexibility of the BSD framework allows us to overcome this limit and to produce a different set of FFTs for each frequency band (e.g. each 10 Hz), without even having to store in files the intermediate product of the analysis. 
An estimation of the analysis gain, for the case of hierarchical all-sky searches, is given in Figure \ref{fig:FFTlength}. In Figure \ref{fig:FFTlength} we show the FFT duration, $T_{FFT}$ which can be used in the two cases (the actual approach, based on the four FFT sets, called the SFDB, and the new BSD approach), as a function of the frequency. The sensitivity improvement for a semi-coherent search (which is $S\propto \frac{1}{\sqrt[4]{T_{FFT}}}$ as described in \cite{Astone2014a}), is given by the fourth root of the ratio of the two durations, $\frac{T^{(BSD)}_{FFT}}{T^{(SFDB)}_{FFT}}$.

In the end, we can consider the collection of BSD files as a database composed by a multitude of sub-databases (the 1 month/10 Hz BSD files) and the new framework a library of basic procedures to efficiently work on a (cleaned) data set\footnote{The framework is based on MATLAB codes, and makes use of the Virgo Rome Snag software \cite{Snag}. The BSD database can obviously be used also with other softwares, provided that the users write their own libraries.}. Our goal in this paper is to concentrate on the application of the new framework to CW analysis, but it is important to consider that we don't see any limit to a possible use of this approach for other searches, like, e.g.,  for long or short duration transients or even for compact binary coalescence signals (see \ref{sec:transients}). In this case, the addition of new libraries,  needed to carry out different analyses, is straightforward.

%

\subsection{The reduced-analytic signal}\label{sec:ana_signal}

In this section, we discuss in more detail the type of data stored in the time series of the BSD. 
The data we store in the BSD is a slightly different version of the classical analytic signal, but it is still a complex-valued time series with no negative frequency components \cite{Gabor1946}.
In general, the real and imaginary parts of any analytic function are related by the Hilbert transform. In particular, if $g(t)$ is a real valued function of $t$, its analytical representation $A(t)$ is given by: 
\begin{equation}
A(t)=g(t)+i\mathcal{H}[g(t)],
\end{equation}
where $\mathcal{H}$ is the Hilbert transform operator.
Since the Fourier ($\mathcal{F}$) and the Hilbert transform are related by  $\mathcal{F} \{\mathcal{H} [g(t)]\}(f)= (-i \mathrm{sign}(f))\cdot \mathcal{F}\{g(t)\}(f)$, the Fourier transform of an analytic signal will be given by:


\begin{equation}\label{eq:fourie_analit}
\mathcal{F}\{A\}(f)= \left\{
                \begin{array}{ll}
				  X(0)\quad \mathrm{if} \,f=0, \qquad   \\
        		  2 X(f)\quad \mathrm{if} \,f>0, \qquad \\
			      0 \quad \mathrm{otherwise},
			    \end{array}
			    \right.	   
\end{equation}
where $X(f)=\mathcal{F}\{g(t)\}(f)$ is the Fourier transform of the  function $g(t)$.  
From the Hermitian symmetry of the Fourier transform, in the analytic representation of a signal the negative frequency components of the Fourier transform can be discarded, with no loss of information, since they are only zeros.

\begin{figure}[h!]
\centering
\includegraphics[scale=0.5,trim=2.8cm 2cm 2.8cm 1.5cm,clip]{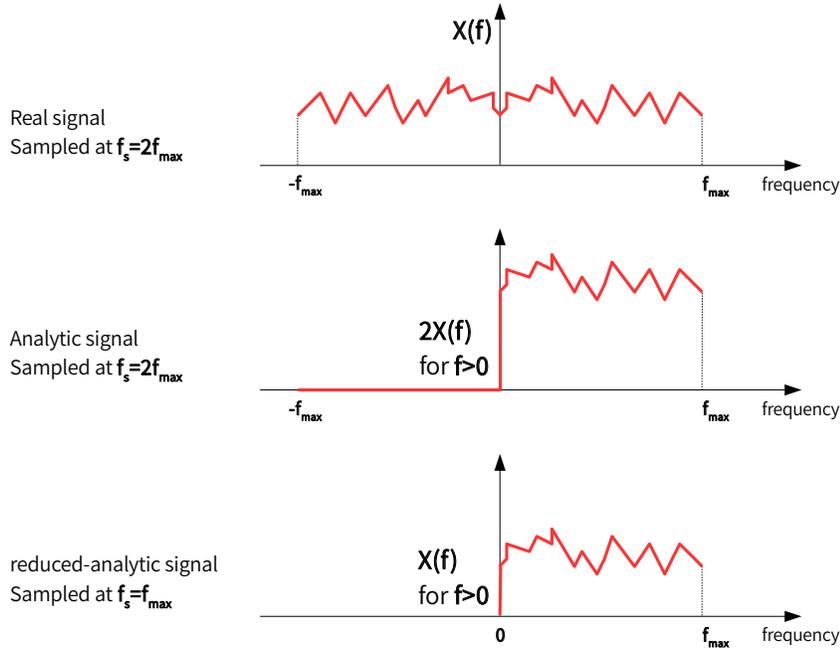}
\caption{Top plot: the Fourier transform of a real valued function would appear with both positive and negative frequency components, which are symmetric around zero due to the Hermitian symmetry. The sampling frequency in this case must be (as required by the Nyquist theorem) at least twice the maximum frequency $f_s=2f_{max}$, in order to avoid aliasing. Middle plot: in 
the  analytic representation of a real-valued function, the negative frequency components of the Fourier transform are zero, while the sampling frequency is the same as in the real-valued case. Bottom plot: our data is built discarding the negative-zero component of a classical analytic signal and it is sampled exactly at the maximum frequency of the positive band (the sampling is then half of the sampling frequency used for the analytic signal). A perfect equivalence exists between the analytic and the reduced-analytic signal. In the construction of our reduced-analytic signal, we decided to keep the same value of the power spectum of the real data, hence the Fourier transform amplitude will be coincident with the real one and half of the classical analytic signal (which is, indeed, twice the real one see Eq. (\ref{eq:fourie_analit})).} \label{fig:analytic}
\end{figure}

From the point of view of signal reconstruction, for a standard analytic signal the sampling frequency is equal to that of the original real valued signal, i.e. at least two times the maximum frequency of the band, as required by the Nyquist theorem. In fact, our data are sampled exactly at the maximum frequency of the band, after shifting the initial frequency to 0 Hz, e.g. at 10 Hz if the data covers a band of 10 Hz, see Figure \ref{fig:analytic}.  
Such signal is called (in this context) \textit{reduced-analytic}, because the sampling frequency is half of the one used for analytic signals. 
However there exists a perfect equivalence between the analytic signal and the reduced-analytic signal: indeed the reduced-analytic signal can be obtained taking only the odd time samples of an analytic signal.
By using appropriate normalization factors a reduced-analytic signal with the following properties can be built:
a) the amplitude (in time) of the complex reduced-analytic signal is half that of the original real data;
b) the power spectrum of the reduced-analytic signal is the same as that of the starting real data. 
A specific MATLAB function, which converts the reduced-analytic signal in its real version using the correct sampling frequency, has been developed to check this equivalence. Since we mostly analyze data in the frequency domain (e.g. using Fourier transforms) rather than in time domain, we decided to set the power spectrum equal to that of the original data.  

\subsection{The BSD heterodyne correction}\label{sec:correction}
As described in Sec. \ref{sec:DAforCW} a CW signal is mainly modulated by the Doppler shift and by the intrinsic spin-down of the source. These modulations need to be properly considered and, when possible, removed, in order to enhance the  Signal-to-Noise Ratio (SNR) and possibly make a detection.
In this section, we concentrate on the application of the method to CW analysis, by describing the procedure used to demodulate the signal from the Doppler and spin-down effects. The method is based on the heterodyne procedure. 
Signal demodulation can be done using different techniques, typically depending on the knowledge of the source parameters, which has an impact on the frequency resolution used for that particular step of the analysis, and clearly on the available computing power.
One standard procedure is based on data re-sampling \cite{Abadie2011}, which exploits the idea that the modulation effects correspond to a time-dependent delay of the received signal. Another technique is based on the heterodyne method, in which the data are multiplied by a complex exponential function that removes the phase modulation as in \cite{Dupuis2005}. 
Data re-sampling has the advantage of being independent from the frequency (for the Doppler), but is typically computationally expensive. In fact, the sampling time needed to properly reconstruct the signal is inversely proportional to the frequency (Nyquist theorem). We remind that a typical CW analysis covers a frequency range from 10 to 2 kHz. Hence, for a search at the highest frequency, data re-sampling might be computationally demanding.
The application of a method based on the heterodyne is particularly indicated for our BSD framework, and it has been implemented with some differences comparing to \cite{Dupuis2005}. Our starting data is complex, already filtered and sub-sampled before applying the heterodyne, while in \cite{Dupuis2005} the time series, which is real and not sub-sampled at the beginning, is multiplied by the complex exponential factor, a low-pass anti-aliasing filter is applied and finally data is re-sampled. In our approach data is ready to be corrected without the need to apply filters or to re-sample, since this is already done by construction. This difference is useful in order to have a more general data framework which can be used also for other searches if needed.

In the following, we assume the frequency and spin-down parameters for a given source at the reference time $t_0$, which we will now call $[f_{0}, \dot{f_{0}}, \ddot{f_{0}},\ldots]$, to be known.
The signal phase shift due to the source spin-down can be written, starting from  Eq. (\ref{eqn:spindown}), as:
\begin{equation}\label{eqn:spindowncorr}
\phi_{sd}(t)=2\pi \int^{t}_{t_0} \left[ \dot{f_0}(t'-t_0) + \frac{1}{2}  \ddot{f_0} (t'-t_0)^{2}+ \ldots \right] dt'.
\end{equation} 
The corresponding phase factor for the Doppler correction is, apart from an irrelevant constant term\footnote{Over long time intervals the integral over $\frac{\vec{p}\cdot\hat{n}}{c}$ is less relevant than the remaining terms, while over small time intervals the spin-down terms can be discarded}:
\begin{equation}
\phi_{d} (t) = 2\pi \int^{t}_{t_0}  f_0(t')\frac{\vec{v}\cdot\hat{n}}{c} dt' \approx \frac{2\pi}{c}  p_{\hat{n}}(t)  f_0(t).
\label{eq:phased}
\end{equation}
where $p_{\hat{n}}(t)$ is the position of the detector in the chosen reference frame, projected along the source sky position $\hat{n}$. The detector position $p_{\hat{n}}(t)$ is easily obtained by interpolating the information stored in the auxiliary data structure (the detector position is originally given at the middle time of each FFT and is referred to the Solar System Barycenter (SSB)).
The total signal phase  correction $\Phi_{corr}(t)=\phi_{d}(t) + \phi_{sd}(t)$, is the sum of the spin-down and the  Doppler contributions.
Heterodyne demodulation is then applied by multiplying the data  by the exponential factor $e^{-i\Phi_{corr}(t)}$:
 \begin{equation} \label{corrected}
y(t)=\left[ h(t)+n(t) \right] e^{-i\Phi_{corr}(t)},
\end{equation}
where $h(t)$ is the strain amplitude of a GW signal in the detector, as given in Eq. \ref{h(t)} for CW signals,  while  $n(t)$ is the detector noise. Once this correction has been applied, a CW signal would become monochromatic,  except  for residual modulations.
These might be present due to inappropriate modeling of the frequency evolution (higher order spin-down terms, source frequency glitches) or to parameter uncertainties like a not perfect estimation of source position parameters (in some cases, source position is estimated during the analysis procedure itself and obviously it is affected by errors, whose entity depends on other parameters, like the Signal-to-Noise Ratio).
The amplitude modulation due to the antenna pattern, mentioned in Sec. \ref{sec:DAforCW}, is different for each detector of the network and can be used to build a detection statistic as well as to estimate signal parameters \cite{Astone2010,Abadie2011}.

\section{Method validation}\label{sec:injections}
In this section we discuss the BSD method validation, by applying the new analysis procedure to LIGO-Virgo data where either ``hardware'' or ``software'' artificial signals, with different parameters, have been added. We call ``hardware'' injections those signals added to the data stream of LIGO-Virgo detectors at the hardware level \cite{Hardwareinj}. 
In any case, in order to have more signals to verify the procedures, it is also typically needed to add them also via ``software'', that is to the data series (either in time or in frequency, depending on the cases).
In the following we describe the results obtained in both cases.

\subsection{Software injections}\label{sec:software}

In order to add CW signals to the collected data set, we have created discrete time series in the standard LIGO-Virgo data format\footnote{The format is called ``gravitational wave frame''. We used the LALAPPS program $\mathtt{lalapps\_Makefakedata\_v4}$ from LALSUITE \cite{LALAPPS} to produce the frames.}.
The frame time series contains only a CW signal. 
The validation procedure described here covers one month of data. We have created BSD files, following all the steps which we have described and which we use for the real data set as well. 
 
\begin{figure}
\centering
\includegraphics[width=0.5\linewidth]{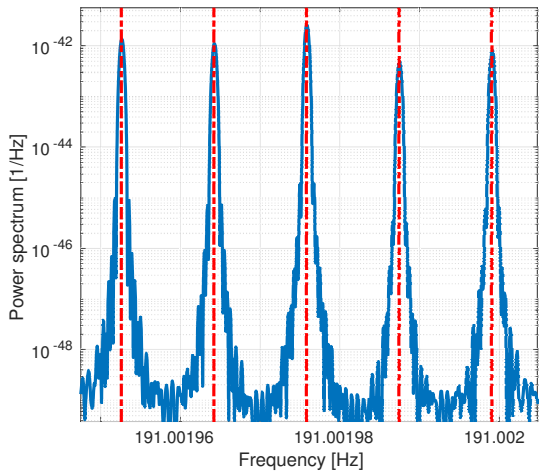}
\caption{(color online) Zoom of the final power spectrum, around the expected injected signal frequency. The sidereal amplitude modulation, remains even after applying the Doppler and spin-down demodulation procedures. Dashed (red) lines identify the theoretical frequencies of the sidereal peaks. In the example, all these effects were removed without significant errors,   as the parameters are completely known and no noise was added.  We notice that the relative peak amplitudes depend on the source location and on the detector position and orientation on the Earth. }
\label{fig:pulsar_frame_8}
\end{figure}

We have first of all verified the very basic properties, i.e. that the reduced-analytic signal used in the BSD is a correct representation of the original frame data and it is coincident with the classical analytic signal. This means that the complex amplitude in the BSD time series is the half of that in the real data, as we expect from the analytic signal theory, and that the spectral amplitude computed with BSDs and with the original data are identical. Details are shown in \ref{sec:ana_signal}.
We have then verified the heterodyne correction procedure implemented within the new BSD framework. Several CW signals, with spin-down and Doppler modulations, have been added to the data and their characteristics after the reconstruction have been studied.
One example, in the absence of noise, is shown in Figure \ref{fig:pulsar_frame_8}. 
The plot shows a zoom of the power spectrum of the signal (whose parameters were chosen identical to those of the hardware injection called ``pulsar 8'' and shown in Table \ref{tabel:inj_param_frame}), after applying Doppler and spin-down demodulations. The  plot shows a zoom around the central expected signal frequency, in order to see the effect of the sidereal amplitude modulation which remains even after the demodulation procedures.
 In the example, all the effects were removed without significant errors, since the parameters were completely known (as in the case of ``targeted searches'') and in this case the signal has been generated without any noise.  

\subsection{Hardware injections}
Validation through the use of ``hardware'' injections has been done in order to fully characterize the method by checking not only our ability to detect CW signals (now obviously embedded in noise) but also to estimate their parameters, which is one of the most delicate part of the procedure. We have used the data of the first Advanced LIGO scientific run, called O1. Injected signals relevant parameters are shown in Table \ref{tabel:inj_param_frame}. We have here used the conventional LIGO-Virgo names for the injections. 
The analysis was done using a single BSD file covering 1 month (from 12-Dec 2015 until the end of O1 run on the 12-Jan 2016) and 10 Hz frequency band containing the signal. Reference time for the parameters is the beginning of the BSD files considered.
\begin{table}
\caption{CW hardware injected signals parameters, referred at the UTC time 12-Dec-2015 00:42:23. Parameters are defined in Sec. \ref{sec:DAforCW} }
\label{tabel:inj_param_frame} 
\begin{indented}
\item[]\begin{tabular}{@{}cccccc}
\br
Injection name & $f_0$ (Hz) & $\dot{f}_0$ (Hz/s) & $(\alpha,\delta)^{\circ}$ &$\cos \iota$&  $\psi$ $(^\circ)$\\ 
\mr
 pulsar 2 & 575.163521 & $-1.37 \times 10^{-13}$ & (215.26,   3.44) & -0.92 & 77.3949\\ 
 pulsar 3 & 108.857159 & $-1.46\times 10^{-17}$  & (178.37, -33.44) & -0.08& 25.4390\\ 
 pulsar 5 &  52.808324 & $-4.03\times 10^{-18}$  & (302.63, -83.84) &  0.46 & -20.85\\ 
 pulsar 8 & 191.001976 & $-8.65\times 10^{-9}$   & (351.39, -33.42) &  0.074 & 9.7673\\ 
\br
\end{tabular}
\end{indented}
\end{table}

Parameter estimation is done by combining the results obtained with the two detectors \cite{Astone2010}. 
In Table \ref{tabel:inj_param} we report the estimated parameters, compared to the true values; $\epsilon_{H_0}$  is the ratio of the estimated over the injected amplitude; $\epsilon_{\eta}$ and $\epsilon_{\psi}$  are the normalized relative errors on $\eta$ and $\psi$. We remind that  the degree of polarization $\eta=-\frac{2\cos \iota}{1+\cos^2 \iota}$ depends on $\iota$  which is the angle between the star rotation axis and the line of sight. We also report the coherence which characterizes the detection reliability (values nearer to one mean a more reliable detection, see \cite{Astone2010}). Results have been compared with those of a standard targeted search pipeline and found in full agreement \cite{Abadie2011}. 
\begin{table}
\caption{\label{tabel:inj_param} Parameters mismatch ($\epsilon_i$) for 1 month analysis and four CW hardware injected signals; the mismatch is computed for $H_0$,  $\eta$ and $\psi$.
$\epsilon_{H_0}$ is the  ratio between the estimated and injected amplitude; $\epsilon_{\eta}$ and $\epsilon_{\psi}$  are respectively the normalized relative errors on $\eta$ and $\psi$. Coherence is a measure of the detection reliability. We notice that the value of $\psi$ recovered for pulsar 2 is badly estimated. This is due to the nearly circular polarization of this hardware injection, which makes $\psi$ ill-defined. Results are in perfect agreement with those obtained with other procedures \cite{Abadie2011}.}
\begin{indented}
\item[]\begin{tabular}{@{}cccccc}
\br
 Injection name & $\epsilon_{H_0}$ & $\epsilon_{\eta}$ & $\epsilon_{\psi}$ & coherence\\
\mr
pulsar 2 & 0.97 & -0.050  & -0.412 & 0.99 \\
pulsar 3 & 1.05 &  0.001  & -0.003 & 0.99\\
pulsar 5 & 0.96 &  0.019  & -0.004 & 0.98 \\
pulsar 8 & 0.96 & -0.005  & -0.009 & 0.99 \\
\br
\end{tabular}
\end{indented}
\end{table} 

\section{Applications}\label{sec:applica}
Some of the CW searches, as already described, are typically computationally constrained, hence a gain in computing power is directly translated into a gain in terms of search sensitivity  \cite{Frasca2005} or into the possibility of enlarging the parameter space searched for. This opens the possibility, for example, to run searches for younger NSs, spinning faster and with an higher spin-down than older NSs, or to lower the analysis thresholds, thus selecting more candidates on which to apply refined ``follow-up'' procedures \cite{Astone2014a}. The possibility of having a flexible and efficient data handling, like the one at the basis of the BSD framework, can make the difference between a nice and interesting new upper limit and a detection.
This stated, the range of possible applications is quite wide and probably limited only by our ability to construct and maintain good libraries for data management, processing (e.g. filtering to enhance the SNR for each given class of signals), analysis, computation of detection confidence.
As the number of detectors in the GW network increases, all these aspects assume a major role. An obvious, but relevant, consideration is that the processing of the data from a network of three detectors is, for example, more computationally demanding than the processing done within a network of two detectors. In these cases, the BSD organization and preprocessing based on cleaning, time and/or frequency extractions, is the right approach to use for very different searches, spanning from very short transients to the long-lasting CW signals, aim of the present paper. 
Going into details for aspects related to CW searches, we have estimated that the analysis of one year of data from three detectors for a single CW source (that is for fixed values of the sky position and rotational parameters) takes about 90 core-minutes on an average processor, which is more than two orders of magnitude better than the old procedure used in \cite{Aasi2014}. The main difference is given by the use of the heterodyne instead of the re-sampling.
 This means, for example, that the search for 200 pulsars, like that described in \cite{Abbott2017o}, could be done on the data of the LIGO-Virgo O3 run, planned to start at the beginning of the year 2019 and to last for presumably about one year, in O(300) core-hours.


The application of the described framework to targeted searches is straightforward, as here cleaning, frequency sub-bands extractions, data demodulation, use of longer FFTs, are at the basis of the search.
The same approach has, however, an immediate utility for the follow-up stage of any kind of semi-coherent CW search. In fact, any of these searches end up with a given number of possible signals, the candidates, and only further and more refined analysis can confirm or reject the candidate.
 In particular, this is done by analyzing the data in a very small region of the parameter space around the candidate. The BSD framework is one of the easiest solution to do this.
 A typical two stages follow-up, for example, where the FFT duration (the coherence time) is increased by a factor of 10, would allow  O($10^5$) follow-ups in a few days, as only few minutes are needed for each follow-up.
 
As already said, cleaning is also a very important part of the procedure and might be used also to study the characteristics of the data. This can be done, for example, analyzing the information obtained during the cleaning process, and stored in the auxiliary BSD data, to search for persistent lines or combs in the spectrum. This is a by-product of the analysis and tools like \emph{NoEMi} \cite{Accadia2012}, a noise mining tool based on CW software, and actually used in the LIGO-Virgo collaboration for detector characterization \cite{Covas2018}. Indeed, the information stored in the auxiliary data might have a great impact in the characterization of GW detectors.

\subsection{Sensitivity estimation for semi-coherent directed searches}\label{sec:sens}
\begin{figure}[h]
\begin{center}
\includegraphics[scale=0.35]{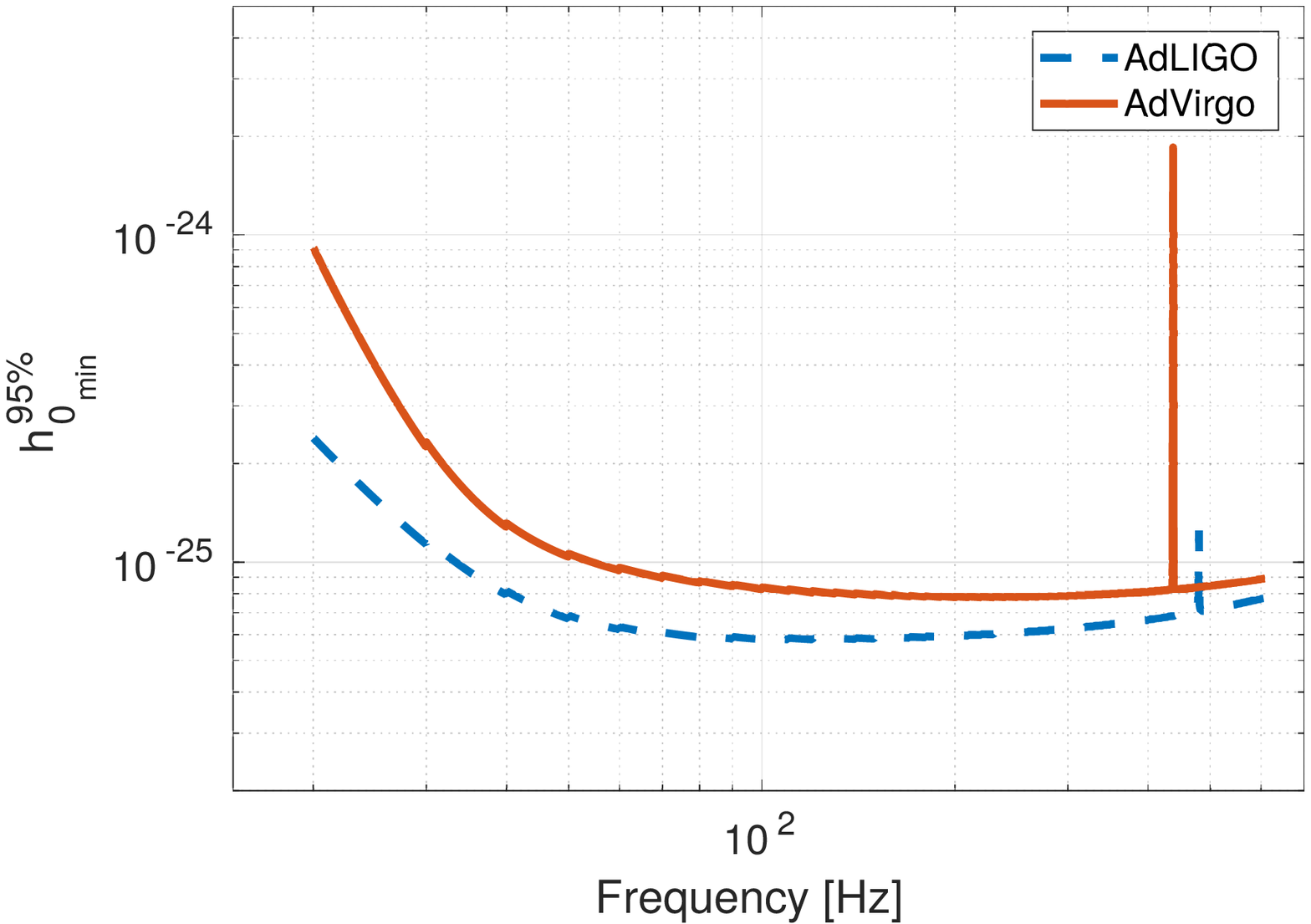}
\includegraphics[scale=0.35]{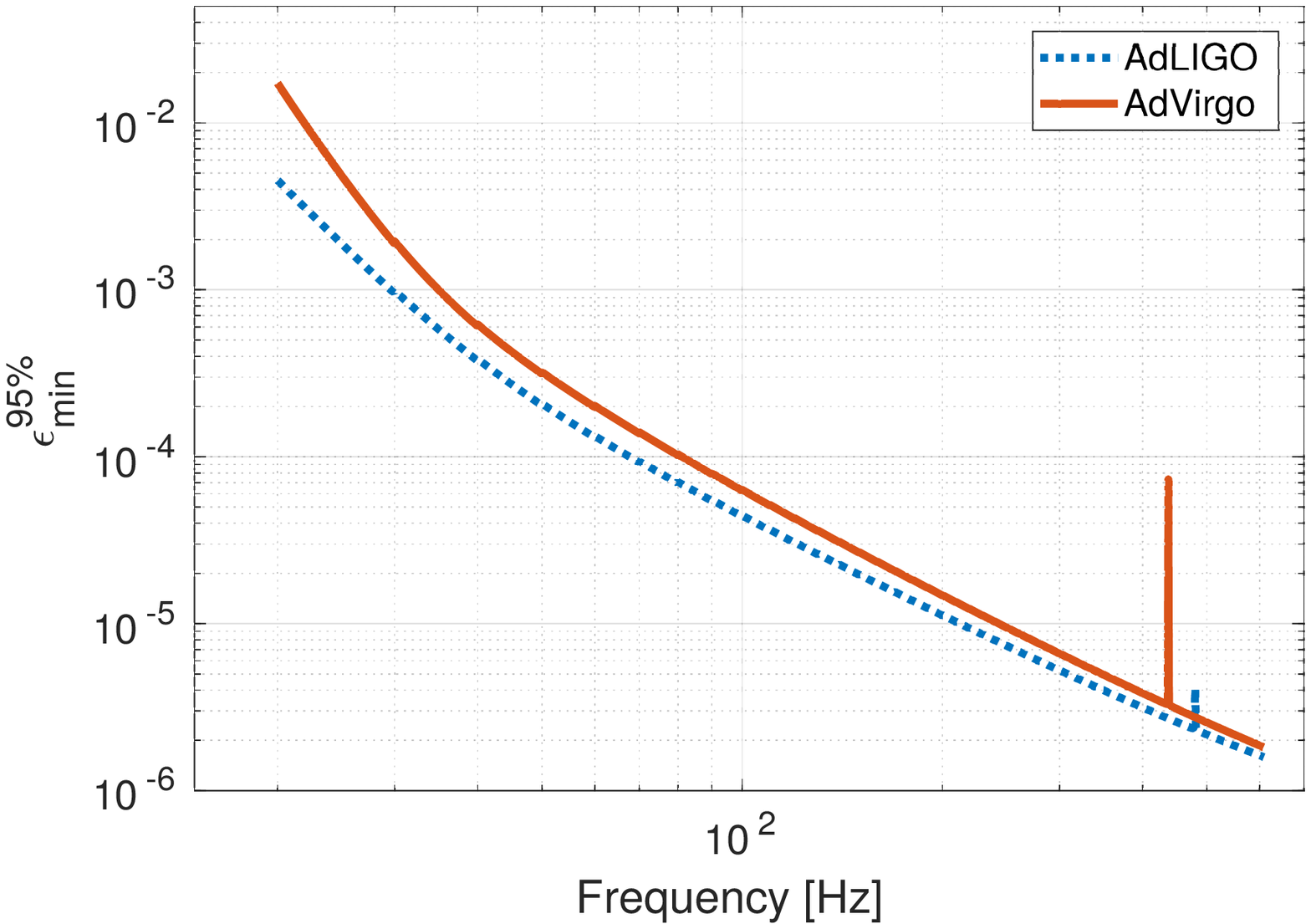}\\
\caption{(color online) Left: Sensitivity estimate for semi-coherent searches, at 95\% C.L. The noise spectra used for the estimate of $h_{0_{min}}$ are those of the Advanced LIGO (dashed line) and Advanced Virgo (solid line) at the design sensitivity for O3 with 6 months of  data.
Right: Minimum detectable ellipticity for a semi-coherent search, pointing to the Galactic Center.  The plot shows the ellipticity of a source in the few inner parsecs of the Galactic Center, which is detectable by the BSD based directed search.}
\label{Fig:sensitivity_ellip}
\end{center}
\end{figure}

As one of the many possible applications of the BSD framework, we have developed a new  CW directed search procedure, whose details will be described in a separate work. It consists in a semi-coherent method for sources where the only known parameter is the sky position (e.g. a search pointing to supernova remnants) or for which a small sky region is assumed to hold several CW sources (e.g. the Galactic Center region). Indeed, as an example, we describe the performances of a semi-coherent directed search based on the BSD-heterodyne framework pointing to the Galactic Center. Another semi-coherent directed search has been performed on two years of data from LIGO’s fifth science run as in \cite{GalCentEaH}, pointing to the Galactic Center.
An interesting number of potential CW sources is located in the few inner parsecs of the Galactic Center \cite{GCpop}. Hence for this search we can consider only a single sky bin, centered at the position of the super-massive black-hole Sagittarius A*. 
With this configuration we can analyze a parameter space of $2.5 \times 10^{10}$ templates, for 6 month of data, opportunely choosing the FFT length for each 10 Hz band. Performances of this search show that the parameter space chosen can be analyzed in less than 20 hours on a workstation Intel(R) Xeon(R) CPU E5-2620 v3 @ 2.40GHz.  
We can provide a theoretical estimation of the search sensitivity for hierarchical semi-coherent searches, using Eq. 67 in section V of \cite{Astone2014a}, which is  valid  under the assumption of Gaussian noise. We can evaluate the sensitivity of the method in terms of the minimum detectable amplitude strain, $h_{0_{min}}$ at 95\% confidence level.
We use the  Advanced LIGO-Virgo detectors design noise spectra, which is expected to be reached during O3 run, as in \cite{Abbott2018,AdVirgo,AdLIGO}. The best sensitivity value is reached around 230 Hz for Virgo and 125 Hz for LIGO,  and are respectively $7.8 \times 10^{-26}$ and $5.9 \times 10^{-26}$. Results are shown in Figure \ref{Fig:sensitivity_ellip}, left.

As typically done in CW searches, the sensitivity $h_{0_{min}}$ can be translated into the minimum ellipticity of potentially detectable
sources. In fact, if we consider a NS spinning around one of its principal axes we have
\begin{equation}
\epsilon=\frac{c^4}{4\pi^2G}\frac{d}{I_{zz}f_0^2}h_{0},
\label{eqn:elliptic}
\end{equation}
where $I_{zz}$ is  the NS moment of inertia with respect to the rotation axis, assumed in the following to be $I_{zz}=10^{38}$ kg m$^2$, as
expected for a standard NS, $d$ is the source distance and $f_0$ is the source GW  frequency. The source distance is $8$ kpc, equal to the Galactic Center distance. The best results, as shown in Figure \ref{Fig:sensitivity_ellip}, right, are obtained in the highest frequency range. The minimum detectable ellipticity for Virgo is $1.8 \times 10^{-6}$ while for LIGO is $1.6 \times 10^{-6}$. The values on the right plot of Figure \ref{Fig:sensitivity_ellip} are comparable to the expected maximum ellipticity of a standard NS and are significantly smaller than the maximum ellipticity foreseen by more exotic equations of state, like quark stars or hybrid stars \cite{McDaniel2013}. 

\section{Conclusions}\label{sec:conclusion}
The detection of gravitational CW signals from isolated and binary spinning NSs is one of the main targets of current and future gravitational wave detectors.
In order to reach this goal, in parallel with detector enhancements, many different algorithms have been developed, with different characteristics in terms of sensitivity and robustness, and hence with different demands in terms of computing power.
In this paper we have presented a novel technique (the BSD framework), that has the goal to enhance the sensitivity of CW searches, maintaining the robustness of the analysis, or to reduce, in some cases substantially, the total computing needs at a fixed sensitivity. This is reached through the use of an efficient data handling procedure which, associated to a  library of functions, gives the possibility to easily perform basic operations on the data, like frequency and or time selections, cleaning, time to frequency (and vice-versa) transformations on the selected data or demodulation.
Some applications have been described here and the sensitivity estimates of a new directed search pipeline, based on BSD,  have been presented. Within this framework is easy to speed up several CW analysis steps, and in particular to critically lower the computational cost of the candidates follow-up, hence increasing the possibility of a detection.
We want to remark that the proposed framework may have an impact also for other GW analysis. Indeed, the low computing cost needed for the files production plus the low storage needs of the collection, enables its portability within the GW community.



\appendix

\section{Data cleaning} \label{sec:cleaning}
BSD data are cleaned using various techniques. First, the short FFTs  database used to build the BSD are vetoed for non-science segments (i.e. data corresponding to time periods when the detector was not locked or not working properly are set to zero) and subject to a cleaning step where big, short duration time disturbances are removed \cite{Astone2005}. A further cleaning step is applied to the BSD data, which again consists in setting to zero large time-domain disturbances which appear after band extraction (and not visible in the starting time series as their power is confined in a small frequency band). The fraction of data put to zero depends on the threshold used for the cleaning. The threshold is taken, after a study based on real detector data, as $\theta_{thr}=10 m_l$, where $m_l$ is evaluated from non zero samples $y_{t_{i;{\tilde{0}}}}:\{y(t_i)\neq 0 \}$ and using the quadratic sum of the median of the real part plus the median of the imaginary part  of the data as:
\begin{equation}
m_l=\sqrt{\mathrm{median}(Re(y_{t_{i;{\tilde{0}}}}))^{2}+\mathrm{median}(Im(y_{t_{i;{\tilde{0}}}}))^{2}}
\label{Eq:ml}
\end{equation}
\begin{figure}[h!]
\centering
\includegraphics[width=0.6\linewidth]{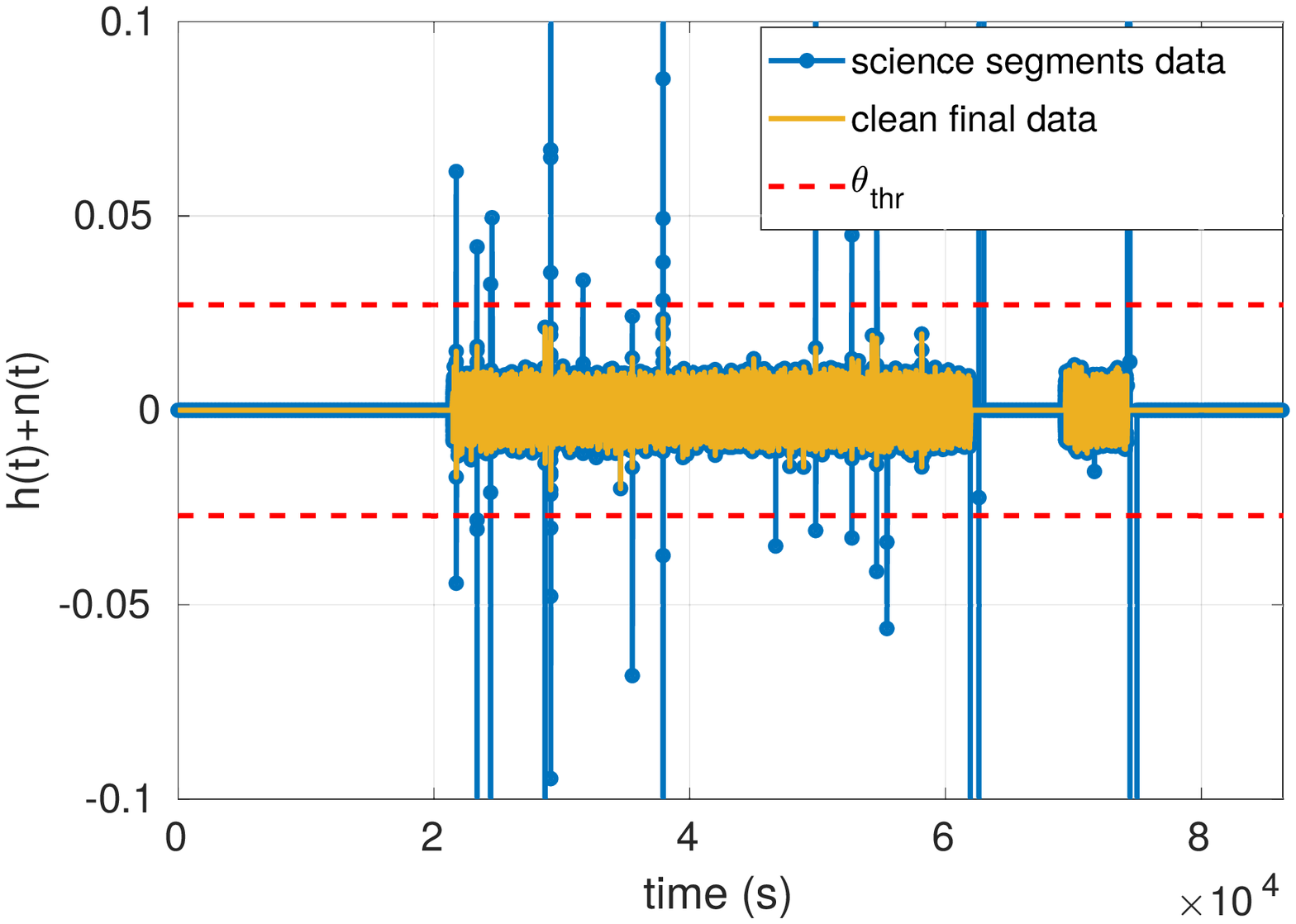}
\caption{Time strain data (in units of $10^{-20}$) before any cleaning is shown in blue. The selection of science data is done directly during SFDB construction (dotted-yellow line). On the BSD time series a further cleaning step is performed to remove residual large time-domain disturbances higher than a given threshold $\theta_{thr}$, computed using Eq. \ref{Eq:ml} (dashed-red line): final cleaned data).}
\label{fig:cleaning}
\end{figure}
As an example, for O1 data the fraction of data excluded with this choice  is around $2 \%$ for Livingston detector and $1 \%$ for Hanford, see Figure \ref{fig:cleaning}. 
This is by purpose a \textit{conservative} cleaning: with this threshold choice we are sure we are not excluding useful data. Furthermore, the threshold is chosen using the median rather than the mean value. In this way big disturbances count as much as normal data without biasing the noise level. An extra cleaning procedure which deletes the more persistent lines and apply a time-frequency filter, built adaptively to the data, can be also applied (see Sec. XIII in \cite{Astone2014a} and \cite{Acernese2009}).
The informations used for the cleaning are stored in the auxiliary time-frequency structure. 
 
\section{The long-band short-period configuration} \label{sec:transients} 
\begin{figure}[h!]
\centering
\includegraphics[width=1\linewidth,trim=0cm 0cm 0cm 1cm,clip]{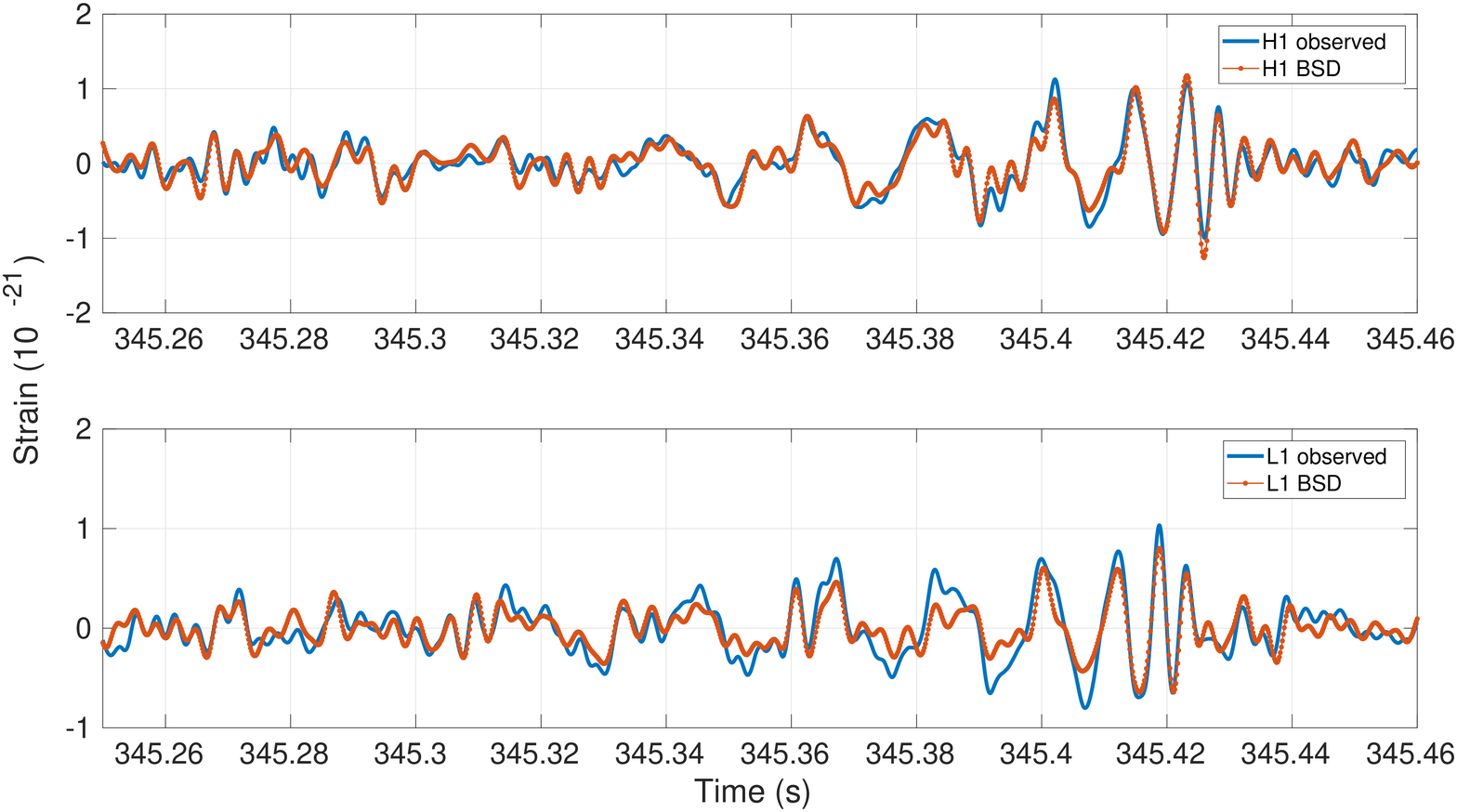}
\caption{(color online). Comparison between the time series of the gravitational wave event GW150914 observed by the LIGO detectors (H1,L1) and the reconstruction of the time series using the BSD procedure for both detectors. Time is relative to 14-September, 2015 at 09:45:00 UTC. The implementation of the filtering procedure used in \cite{Abbott2016} is different from the one used for BSD data, this explain the main fluctuations between the two time series  the same detector. The resulting BSD time series is obtained extracting a given frequency band. In addition to this first selection, a white filter in the frequency domain has been applied. The final time series is obtained with an inverse Fourier transform. 
Top panel: the solid (blue) line shows the time series  as observed by the LIGO Hanford (H1) detector, while the BSD data is the (orange) solid-dotted line. 
Low panel: the solid (blue) line shows the time series as observed by the LIGO Livingston (L1) detector, while the  BSD data is the (orange) solid-dotted line. }
\label{fig:event}
\end{figure}

As mentioned in \ref{sec:method}, this data framework can be used for the search of transient signals, since it is possible to manage the BSD files in order to create a new file covering a large frequency band and a short time period, which is the typical case of the gravitational wave signals detected so far \cite{Abbott2016,Abbott2017n,Abbott2017a,Abbott2017k,Abbott2017,Abbott2016f,Abbott2016a}. 
The long-band short-period configuration is obtained by concatenating in the Fourier domain, the FFTs of all the adjacent frequency bands considered. The time series is then obtained doing the inverse FFT of the concatenated bands. This procedure has been used for the extraction of data around the GW event GW150914, taking a frequency band of [30-320] Hz and a time span of 10 minutes, from 09:45:00 to 09:55:00 UTC on the 14-September 2015. The extracted time series for Hanford and Livingston detectors can be seen in Figure \ref{fig:event}, where a signal is clearly visible after applying a whitening filter on the data of the new BSD. The extracted and filtered time series is compatible with the one observed in the two detectors (see top row in Figure 1 of \cite{Abbott2016}). The main difference between the  BSD time series and the detector ones is due to the independent filtering procedure used. More information about the original detector time series used for Figure \ref{fig:event},  can be found in \cite{Abbott2016,LOSC}.

\section*{References}

\bibliographystyle{unsrt}
\bibliography{mybib}

\begin{thebibliography}{10}

\bibitem{Abbott2016}
Abbott~B.P. et~al.
\newblock {Observation of gravitational waves from a binary black hole merger}.
\newblock {\em Phys. Rev. Lett.}, 116(6):061102, 2016.

\bibitem{Abbott2017n}
Abbott~B.P. et~al.
\newblock {GW170817: Observation of Gravitational Waves from a Binary Neutron
  Star Inspiral}.
\newblock {\em Phys. Rev. Lett.}, 119(16):161101, 2017.

\bibitem{Abbott2017a}
Abbott~B.P. et~al.
\newblock {GW170608: Observation of a 19 Solar-mass Binary Black Hole
  Coalescence}.
\newblock {\em The Astrophysical Journal Letters}, 851(2):L35, 2017.

\bibitem{Abbott2017k}
Abbott~B.P. et~al.
\newblock {GW170814: A Three-Detector Observation of Gravitational Waves from a
  Binary Black Hole Coalescence}.
\newblock {\em Phys. Rev. Lett.}, 119:141101, 2017.

\bibitem{Abbott2017}
Abbott~B.P. et~al.
\newblock {GW170104: Observation of a 50-Solar-Mass Binary Black Hole
  Coalescence at Redshift 0.2}.
\newblock {\em Phys. Rev. Lett.}, 118(22):221101, 2017.

\bibitem{Abbott2016f}
Abbott~B.P. et~al.
\newblock {Binary Black Hole Mergers in the First Advanced LIGO Observing Run}.
\newblock {\em Physical Review X}, 6(4):041015, 2016.

\bibitem{Abbott2016a}
Abbott~B.P. et~al.
\newblock {GW151226: Observation of Gravitational Waves from a 22-Solar-Mass
  Binary Black Hole Coalescence}.
\newblock {\em Phys. Rev. Lett.}, 116(24):241103, 2016.

\bibitem{Lasky2015}
Lasky P.D.
\newblock {Gravitational Waves from Neutron Stars: A Review}.
\newblock {\em Publ. Astron. Soc. of Aust.}, 32:e034, 2015.

\bibitem{Glampe}
Glampedakis K. and Gualtieri L.
\newblock {Gravitational waves from single neutron stars: an advanced detector
  era survey}.
\newblock {\em ArXiv e-prints}, September 2017.

\bibitem{Astone2010}
Astone P., D'Antonio S., Frasca S., and Palomba C.
\newblock {A method for detection of known sources of continuous gravitational
  wave signals in non-stationary data}.
\newblock {\em Classical and Quantum Gravity}, 27(19):194016, 2010.

\bibitem{Astone2014c}
Astone P., Colla A., D'Antonio S., Frasca S., Palomba C., and Serafinelli R.
\newblock {Method for narrow-band search of continuous gravitational wave
  signals}.
\newblock {\em Phys. Rev. D}, 89:062008, 2014.

\bibitem{Mastrogiovanni2017}
Mastrogiovanni S., Astone P., D'Antonio S., Frasca S., Intini G., Leaci P.,
  Miller A., Palomba C., Piccinni~O. J., and Singhal A.
\newblock {An improved algorithm for narrow-band searches of continuous
  gravitational waves}.
\newblock {\em Classical and Quantum Gravity}, 34(13):135007, 2017.

\bibitem{GalCentEaH}
Aasi~J. et~al.
\newblock {Directed search for continuous gravitational waves from the Galactic
  center}.
\newblock {\em Phys. Rev. D}, 88:102002, Nov 2013.

\bibitem{Dergachev2010}
Dergachev V.
\newblock {On blind searches for noise dominated signals: a loosely coherent
  approach}.
\newblock {\em Classical and Quantum Gravity}, 27(20):205017, 2010.

\bibitem{Astone2014a}
Astone P., Colla A., D'Antonio S., Frasca S., and Palomba C.
\newblock {Method for all-sky searches of continuous gravitational wave signals
  using the frequency-Hough transform}.
\newblock {\em Phys. Rev. D}, 90:042002, 2014.

\bibitem{Abadie2011}
Abadie~J. et~al.
\newblock {Beating the Spin-down Limit on Gravitational Wave Emission from the
  Vela Pulsar}.
\newblock {\em The Astrophysical Journal}, 737(2):93, 2011.

\bibitem{Aasi2014}
Aasi~J. et~al.
\newblock {Gravitational Waves from Known Pulsars: Results from the Initial
  Detector Era}.
\newblock {\em The Astrophysical Journal}, 785(2):119, 2014.

\bibitem{Abbott2017o}
Abbott~B.P. et~al.
\newblock {First Search for Gravitational Waves from Known Pulsars with
  Advanced LIGO}.
\newblock {\em The Astrophysical Journal}, 839(1):12, 2017.

\bibitem{Abbott2017b_narrow}
Abbott~B.P. et~al.
\newblock {First narrow-band search for continuous gravitational waves from
  known pulsars in advanced detector data}.
\newblock {\em Phys. Rev. D}, 96:122006, Dec 2017.

\bibitem{Aasi2015}
Aasi~J. et~al.
\newblock {Narrow-band search of continuous gravitational-wave signals from
  Crab and Vela pulsars in Virgo VSR4 data}.
\newblock {\em Phys. Rev. D}, 91:022004, 2015.

\bibitem{Abbott2017Sco2}
Abbott~B.P. et~al.
\newblock {Upper Limits on Gravitational Waves from Scorpius X-1 from a
  Model-based Cross-correlation Search in Advanced LIGO Data}.
\newblock {\em The Astrophysical Journal}, 847(1):47, 2017.

\bibitem{Aasi2014a}
Aasi~J. et~al.
\newblock {Searches for Continuous Gravitational Waves From Nine Young
  Supernova Remnants}.
\newblock {\em The Astrophysical Journal}, 813(1):39, 2015.

\bibitem{Aasi2016b}
Aasi~J. et~al.
\newblock {First low frequency all-sky search for continuous gravitational wave
  signals}.
\newblock {\em Phys. Rev. D}, 93:042007, 2016.

\bibitem{Abbott2017w}
Abbott~B.P. et~al.
\newblock {All-sky search for periodic gravitational waves in the O1 LIGO
  data}.
\newblock {\em Phys. Rev. D}, 96:062002, 2017.

\bibitem{Riles2017}
Riles K.
\newblock Recent searches for continuous gravitational waves.
\newblock {\em Modern Physics Letters A}, 32(39):1730035, 2017.

\bibitem{Palomba2017}
Palomba C.
\newblock {The search for continuous gravitational waves with LIGO and Virgo
  detectors}.
\newblock {\em Eur. Phys. Soc. Conf. High Energy Phys.}, 5, 2017.

\bibitem{Jaranowski1998}
Jaranowski P., Kr{\'{o}}lak A., and Schutz B.F.
\newblock {Data analysis of gravitational-wave signals from spinning neutron
  stars: The signal and its detection}.
\newblock {\em Phys. Rev. D}, 58:063001, 1998.

\bibitem{Leaci2017}
Leaci P., Astone P., D'Antonio S., Frasca S., Palomba C., Piccinni O., and
  Mastrogiovanni S.
\newblock {Novel directed search strategy to detect continuous gravitational
  waves from neutron stars in low- and high-eccentricity binary systems}.
\newblock {\em Phys. Rev. D}, 95:122001, Jun 2017.

\bibitem{ShapiroEinsteindelay}
D.C. {Backer} and R.W. {Hellings}.
\newblock {Pulsar timing and general relativity}.
\newblock {\em Annual Review of Astronomy and Astrophysics}, 24:537--575, 1986.

\bibitem{Dupuis2005}
R\'ejean~J. Dupuis and Graham Woan.
\newblock Bayesian estimation of pulsar parameters from gravitational wave
  data.
\newblock {\em Phys. Rev. D}, 72:102002, Nov 2005.

\bibitem{Keitel2018}
David Keitel and Gregory Ashton.
\newblock {Faster search for long gravitational-wave transients: GPU
  implementation of the transient {$ \newcommand{\F}{\mathcal{F}}\boldsymbol{
  \F}$} -statistic}.
\newblock {\em Classical and Quantum Gravity}, 35(20):205003, 2018.

\bibitem{Astone2005}
Astone P., S.~Frasca, and Palomba C.
\newblock {The short FFT database and the peak map for the hierarchical search
  of periodic sources}.
\newblock {\em Classical and Quantum Gravity}, 22:S1197, 2005.

\bibitem{Snag}
{Snag: Signal and noise for gravitational antennas }.
\newblock
  {https://web.infn.it/VirgoRoma/index.php/en/research-eng/data-analysis-eng/snag-eng-page}.

\bibitem{Gabor1946}
Gabor D.
\newblock {Theory of communication.}
\newblock {\em J Inst. Electr. Eng.}, 93(111):429--441, 1946.

\bibitem{Hardwareinj}
Biwer C. and et~al.
\newblock {Validating gravitational-wave detections: The Advanced LIGO hardware
  injection system}.
\newblock {\em Phys. Rev. D}, 95:062002, Mar 2017.

\bibitem{LALAPPS}
{LIGO Scientific Collaboration. Lal/lalapps software suite.}
\newblock {http://www.lsc-group.phys.uwm.edu/daswg/projects/lalsuite.html}.

\bibitem{Frasca2005}
{Frasca S. and Astone P. and Palomba C.}
\newblock Evaluation of sensitivity and computing power for the virgo
  hierarchical search for periodic sources.
\newblock {\em Classical and Quantum Gravity}, 22(18):S1013, 2005.

\bibitem{Accadia2012}
T.~et~al. Accadia.
\newblock {The NoEMi (Noise Frequency Event Miner) framework}.
\newblock {\em Journal of Physics: Conference Series}, 363(1):012037, 2012.

\bibitem{Covas2018}
Covas P.~B. et~al.
\newblock {Identification and mitigation of narrow spectral artifacts that
  degrade searches for persistent gravitational waves in the first two
  observing runs of Advanced LIGO}.
\newblock {\em Phys. Rev. D}, 97:082002, Apr 2018.

\bibitem{GCpop}
Chennamangalam J. and Lorimer D.R.
\newblock {The Galactic Centre pulsar population}.
\newblock {\em Monthly Notices of the Royal Astronomical Society: Letters},
  440(1):L86--L90, 2014.

\bibitem{Abbott2018}
Abbott~B. P. and et~al.
\newblock {Prospects for observing and localizing gravitational-wave transients
  with Advanced LIGO, Advanced Virgo and KAGRA}.
\newblock {\em Living Reviews in Relativity}, 21(1):3, Apr 2018.

\bibitem{AdVirgo}
Acernese~F. et~al.
\newblock {Advanced Virgo: a second-generation interferometric gravitational
  wave detector}.
\newblock {\em Classical and Quantum Gravity}, 32(2):024001, 2015.

\bibitem{AdLIGO}
Aasi~J. et~al.
\newblock {Advanced LIGO}.
\newblock {\em Classical and Quantum Gravity}, 32(7):074001, 2015.

\bibitem{McDaniel2013}
Johnson-McDaniel N.K. and Owen B.J.
\newblock Maximum elastic deformations of relativistic stars.
\newblock {\em Phys. Rev. D}, 88:044004, Aug 2013.

\bibitem{Acernese2009}
Acernese~F. et~al.
\newblock {Cleaning the Virgo sampled data for the search of periodic sources
  of gravitational waves}.
\newblock {\em Classical and Quantum Gravity}, 26(20):204002, 2009.

\bibitem{LOSC}
{GWOSC: Gravitational Wave Open Science Center}.
\newblock {https://www.gw-openscience.org/events/GW150914/}.

\end{thebibliography}

\end{document}